\documentstyle[psfig]{l-aa}
\newhelp\stablestylehelp{You must choose a style between 0 and 3.}%
\newhelp\stablelinehelp{You should not use special hrules when stretching
a table.}%
\newhelp\stablesmultiplehelp{You have tried to place an S-Table inside another
S-Table.  I would recommend not going on.}%
\newdimen\stablesthinline
\newdimen\stablesthickline
\newif\ifstablesborderthin
\stablesborderthinfalse
\newif\ifstablesinternalthin
\stablesinternalthintrue
\newif\ifstablesomit
\newif\ifstablemode
\newif\ifstablesright
\stablesrightfalse
\newdimen\stablesbaselineskip
\newdimen\stableslineskip
\newdimen\stableslineskiplimit
\newcount\stablesmode
\newcount\stableslines
\newcount\stablestemp
\newcount\stablescount
\newcount\stableslinet
\newcount\stablestyle
\def\stablesleft{\quad\hfil}%
\def\stablesright{\hfil\quad}%
\catcode`\|=\active%
\newcount\stablestrutsize
\newbox\stablestrutbox
\setbox\stablestrutbox=\hbox{\vrule height10pt depth5pt width0pt}
\def\stablestrut{\relax\ifmmode%
                         \copy\stablestrutbox%
                       \else%
                         \unhcopy\stablestrutbox%
                       \fi}%
\newdimen\stablesborderwidth
\newdimen\stablesinternalwidth
\newdimen\stablesdummy
\newcount\stablesdummyc
\newif\ifstablesin
\stablesinfalse
\def\begintable{\stablestart%
  \stablemodetrue%
  \stablesadj%
  \halign%
  \stablesdef}%
\def\stablesadj{%
  \ifcase\stablestyle%
    \hbox to \hsize\bgroup\hss\vbox\bgroup%
  \or%
    \hbox to \hsize\bgroup\vbox\bgroup%
  \or%
    \hbox to \hsize\bgroup\hss\vbox\bgroup%
  \or%
    \hbox\bgroup\vbox\bgroup%
  \else%
    \errhelp=\stablestylehelp%
    \errmessage{Invalid style selected, using default}%
    \hbox to \hsize\bgroup\hss\vbox\bgroup%
  \fi}%
\def\stablesend{\egroup%
  \ifcase\stablestyle%
    \hss\egroup%
  \or%
    \hss\egroup%
  \or%
    \egroup%
  \or%
    \egroup%
  \else%
    \hss\egroup%
  \fi}%
\def\stablestart{%
  \ifstablesin%
    \errhelp=\stablesmultiplehelp%
    \errmessage{An S-Table cannot be placed within an S-Table!}%
  \fi
  \global\stablesintrue%
  \global\advance\stablescount by 1%
  \message{<S-Tables Generating Table \number\stablescount}%
  \begingroup%
  \stablestrutsize=\ht\stablestrutbox%
  \advance\stablestrutsize by \dp\stablestrutbox%
  \ifstablesborderthin%
    \stablesborderwidth=\stablesthinline%
  \else%
    \stablesborderwidth=\stablesthickline%
  \fi%
  \ifstablesinternalthin%
    \stablesinternalwidth=\stablesthinline%
  \else%
    \stablesinternalwidth=\stablesthickline%
  \fi%
  \tabskip=0pt%
  \stablesbaselineskip=\baselineskip%
  \stableslineskip=\lineskip%
  \stableslineskiplimit=\lineskiplimit%
  \offinterlineskip%
  \def\borderrule{\vrule width \stablesborderwidth}%
  \def\internalrule{\vrule width \stablesinternalwidth}%
  \def\thinline{\noalign{\hrule height \stablesthinline}}%
  \def\thickline{\noalign{\hrule height \stablesthickline}}%
  \def\trule{\omit\leaders\hrule height \stablesthinline\hfill}%
  \def\ttrule{\omit\leaders\hrule height \stablesthickline\hfill}%
  \def\tttrule##1{\omit\leaders\hrule height ##1\hfill}%
  \def\stablesel{&\omit\global\stablesmode=0%
    \global\advance\stableslines by 1\borderrule\hfil\cr}%
  \def\el{\stablesel&}%
  \def\elt{\stablesel\thinline&}%
  \def\eltt{\stablesel\thickline&}%
  \def\elttt##1{\stablesel\noalign{\hrule height ##1}&}%
  \def\elspec{&\omit\hfil\borderrule\cr\omit\borderrule&%
              \ifstablemode%
              \else%
                \errhelp=\stablelinehelp%
                \errmessage{Special ruling will not display properly}%
              \fi}%
  \def\stmultispan##1{\mscount=##1 \loop\ifnum\mscount>3 \stspan\repeat}%
  \def\stspan{\span\omit \advance\mscount by -1}%
  \def\multicolumn##1{\omit\multiply\stablestemp by ##1%
     \stmultispan{\stablestemp}%
     \advance\stablesmode by ##1%
     \advance\stablesmode by -1%
     \stablestemp=3}%
  \def\multirow##1{\stablesdummyc=##1\parindent=0pt\setbox0\hbox\bgroup%
    \aftergroup\emultirow\let\temp=}
  \def\emultirow{\setbox1\vbox to\stablesdummyc\stablestrutsize%
    {\hsize\wd0\vfil\box0\vfil}%
    \ht1=\ht\stablestrutbox%
    \dp1=\dp\stablestrutbox%
    \box1}%
  \def\stpar##1{\vtop\bgroup\hsize ##1%
     \baselineskip=\stablesbaselineskip%
     \lineskip=\stableslineskip%
     \lineskiplimit=\stableslineskiplimit\bgroup\aftergroup\estpar\let\temp=}%
  \def\estpar{\vskip 6pt\egroup}%
  \def\stparrow##1##2{\stablesdummy=##2%
     \setbox0=\vtop to ##1\stablestrutsize\bgroup%
     \hsize\stablesdummy%
     \baselineskip=\stablesbaselineskip%
     \lineskip=\stableslineskip%
     \lineskiplimit=\stableslineskiplimit%
     \bgroup\vfil\aftergroup\estparrow%
     \let\temp=}%
  \def\estparrow{\vfil\egroup%
     \ht0=\ht\stablestrutbox%
     \dp0=\dp\stablestrutbox%
     \wd0=\stablesdummy%
     \box0}%
  \def|{\global\advance\stablesmode by 1&&&}%
  \def\|{\global\advance\stablesmode by 1&\omit\vrule width 0pt%
         \hfil&&}%
  \def\vt{\global\advance\stablesmode by 1&\omit\vrule width \stablesthinline%
          \hfil&&}%
  \def\vtt{\global\advance\stablesmode by 1&\omit\vrule width \stablesthickline%
          \hfil&&}%
  \def\vttt##1{\global\advance\stablesmode by 1&\omit\vrule width ##1%
          \hfil&&}%
  \def\vtr{\global\advance\stablesmode by 1&\omit\hfil\vrule width%
           \stablesthinline&&}%
  \def\vttr{\global\advance\stablesmode by 1&\omit\hfil\vrule width%
            \stablesthickline&&}%
  \def\vtttr##1{\global\advance\stablesmode by 1&\omit\hfil\vrule width ##1&&}%
  \stableslines=0%
  \stablesomitfalse}
\def\stablesdef{\bgroup\stablestrut\borderrule##\tabskip=0pt plus 1fil%
  &\stablesleft##\stablesright%
  &##\ifstablesright\hfill\fi\internalrule\ifstablesright\else\hfill\fi%
  \tabskip 0pt&&##\hfil\tabskip=0pt plus 1fil%
  &\stablesleft##\stablesright%
  &##\ifstablesright\hfill\fi\internalrule\ifstablesright\else\hfill\fi%
  \tabskip=0pt\cr%
  \ifstablesborderthin%
    \thinline%
  \else%
    \thickline%
  \fi&%
}%
\def\endtable{\advance\stableslines by 1\advance\stablesmode by 1%
   \message{- Rows: \number\stableslines, Columns:  \number\stablesmode>}%
   \stablesel%
   \ifstablesborderthin%
     \thinline%
   \else%
     \thickline%
   \fi%
   \egroup\stablesend%
\endgroup%
\global\stablesinfalse}
\newcommand{\be}{\begin{equation}}
\newcommand{\en}{\end{equation}}
\begin{document}
\def\ltsima{$\; \buildrel < \over \sim \;$}
\def\lsim{\lower.5ex\hbox{\ltsima}}
\def\gtsima{$\; \buildrel > \over \sim \;$}
\def\gsim{\lower.5ex\hbox{\gtsima}}
\def\spose#1{\hbox to 0pt{#1\hss}}
\def\approxlt{\mathrel{\spose{\lower 3pt\hbox{$\sim$}}
        \raise 2.0pt\hbox{$<$}}}
\def\approxgt{\mathrel{\spose{\lower 3pt\hbox{$\sim$}}
        \raise 2.0pt\hbox{$>$}}}
\def\deg {^\circ}
\def\mdot {\dot M}
\def\kms {~km~s$^{-1}$}
\def\gs {~g~s$^{-1}$}
\def\ergs {~erg~s$^{-1}$}
\def\cmtre {~cm$^{-3}$}\def\nupa{\vfill\eject\noindent}
\def\der#1#2{{d #1 \over d #2}}
\def\l#1{\lambda_{#1}}
\def\grb{$\gamma$-ray burst}
\def\grbs{$\gamma$-ray bursts}
\def\rosat{{\sl ROSAT} }
\def\cmdue {~cm$^{-2}$}
\def\gcm {~g~cm$^{-3}$}
\def\rsole{~R_{\odot}}
\def\msole{~M_{\odot}}
\def\aa #1 #2 {A\&A, {#1}, #2}
\def\mon #1 #2 {MNRAS, {#1}, #2}
\def\apj #1 #2 {ApJ, {#1}, #2}
\def\nat #1 #2 {Nature, {#1}, #2}
\def\pasj #1 #2 {PASJ, {#1}, #2}
\newfont{\mc}{cmcsc10 scaled\magstep2}
\newfont{\cmc}{cmcsc10 scaled\magstep1}
\newcommand{\bc}{\begin{center}}
\newcommand{\ec}{\end{center}}

\title{Is RXJ1856.5--3754 an old neutron star?}
\author{S.~Campana\inst{1,\,2}, S. Mereghetti\inst{3} \& L. Sidoli\inst{3}}

\institute{
{Osservatorio Astronomico di Brera, Via Bianchi 46, I-22055
Merate (Lc), Italy; \\ e-mail: campana@merate.mi.astro.it}
\and
{Affiliated to I.C.R.A.}
\and
{Istituto di Fisica Cosmica del C.N.R., Via Bassini 15, I-20133 Milano,
Italy; \\ e-mail: (sandro, sidoli)@ifctr.mi.cnr.it}
}

\maketitle
\label{sampout}

\begin{abstract}

An unusual X--ray source, RXJ1856.5--3754, has recently been
discovered with ROSAT: its spectrum resembles that of supersoft sources,
but the very high X--ray to optical flux ratio excludes the presence of a
companion star, pointing to an isolated compact object. 
It has been proposed that RXJ1856.5--3754 is an old, isolated neutron
star accreting from the interstellar medium. Here we present a reanalysis
of the ROSAT data. The HRI observation reveals an attitude
reconstruction problem, resulting in a larger error box than previously
reported. Deeper optical observations allow us to reveal a few optical
counterparts inside the new error box, but the X--ray to optical flux
ratio is still very high ($\gsim 20$) and none of the candidates has
peculiar colors.  

\end{abstract}
 
\section{Introduction}

Though only about 700 radio pulsars are presently known, the Galaxy
contains a large population of old neutron stars (ONSs).
Estimates of the neutron star birth rate, integrated over the
Galaxy lifetime, predict that there are about $10^8-10^9$ ONSs.
These elusive objects could be discovered through the
detection of radiation powered by accretion from the interstellar medium
(Ostriker, Rees \& Silk 1970). Since this emission is expected to peak
in the UV and soft X--ray range, the advent of the ROSAT satellite
stimulated much theoretical work on the subject (Treves \& Colpi 1991;
Blaes \& Madau 1993; Colpi, Campana \& Treves 1993), as well as
observational searches which recently led to the proposal of a few
candidate ONSs (Belloni, Zampieri \& Campana 1996; Stocke et al. 1995;
Walter, Wolk \& Neuh\"auser 1996).

A large number of isolated neutron stars still
active as radio pulsars are presently undiscovered simply due to beaming
effects. The few pulsars seen above 100 MeV indicate that the
$\gamma-$ray beam is probably larger than the radio one and, as clearly
shown by Geminga, radio-quiet pulsars are now a reality
(Bignami \& Caraveo 1996). 
On the other hand, the thermal X--ray emission
from these objects is practically visible over 4$\,\pi$, as can be
inferred from the small pulsed fractions of the ROSAT light curves and
from the spectra consistent with emission from the whole (or a very large
fraction of the) neutron star surface (\"Ogelman 1995).
Thus it is not surprising that other ``radio-quiet" neutron star
candidates have recently been found with ROSAT (Mereghetti, Bignami \&
Caraveo 1996; Petre, Becker \& Winkler 1996). In the lack of a
detectable periodicity, the main arguments supporting their neutron
star nature are the location at the center of supernova remnants and
their high X--ray to optical flux ratio ($F_X/F_{opt}$), which rules out
other known classes of X--ray sources.

Walter et al. (1996) have recently reported the discovery of an
unusual source, RXJ1856.5--3754, which they interpreted as a nearby
ONS powered by accretion from the interstellar medium. This
source is the brightest ONS candidate reported
so far. Its main properties (very high $F_X/F_{opt}$,
ultra soft spectrum with blackbody temperature, lack of
variability) have already been reported by Walter et al. (1996). However,
the importance of this source
prompted us to present here a more detailed analysis of the X--ray data
and a more critical interpretation of the results. 

\begin{figure}[!th]
\centerline{\psfig{figure=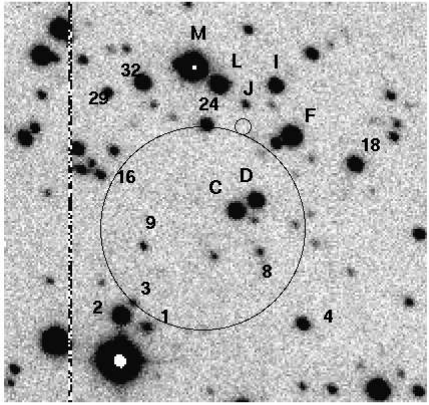,width=7cm} {\hfil}}
\caption{Image of $1.4'\times1.4'$ RXJ1856.5--3754 field ($i$ band).
The center of the larger error box is the PSPC position after the
boresight correction. The smaller circle represents the error circle
derived by Walter et al. (1996). North is to the top and East to the left.}
\end{figure}

\section{The position of RXJ1856.5--3754}

A ROSAT Position Sensitive Proportional Counter (PSPC) observation of 
RXJ1856.5--3754
was carried out during a survey of the RCrA molecular cloud, on October 
11-16, 1992
followed by a deeper ROSAT High Resolution Imager (HRI) pointing performed on
October 7--8, 1994. 

In the 6332 s PSPC observation, RXJ1856.5--3754 was detected at an
off-axis angle of $14.5'$, still in the central region within the
window support rib. 
Due to the large number of counts detected from RXJ1856.5--3754, the
error on its position is dominated by possible systematic uncertainties,
rather than counting statistics. To evaluate this uncertainty 
proceeded
we applied a maximum likelihood detection algorithm provided with the
MIDAS/EXSAS software to the inner ($40'$ diameter) PSPC image corresponding to the
0.4--2.4 keV energy range.  
Ten sources were detected above a threshold
corresponding to a chance occurrence probability of $5\times 10^{-5}$.
Due to the crowding of this low galactic latitude field, their
identification with objects visible on the digitized sky plates is quite
problematic.
Seven of these sources have relatively bright (${\rm V} \gsim 15$)
stars of the HST Guide Star Catalogue (GSC) within their error boxes.
Based on the resulting $F_X/F_{opt}$, we considered these stars as plausible 
identifications, and used them to calibrate, with a least square fit, 
the X--ray reference frame. This resulted in the best fit position 
18h 56m 36s, $-37\deg 54' 52''$ indicated by the error circle in Figure 1. 
The radius of $20''$ corresponds to the average residuals of the fit.

\begin{figure}[!th]
\centerline{\psfig{figure=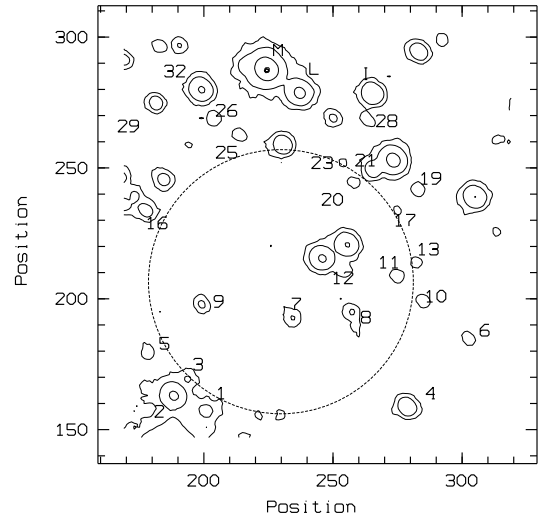,width=9cm} {\hfil}}
\caption{Contour plot of the central region of Figure 1.
Many faint sources can be observed in this figure inside the error box.}
\end{figure}

Walter et al. (1996) derived a much smaller error circle based on the
17800 s HRI pointed observation. 
However, a careful analysis of the radial and azimuthal distribution of
the source counts in the HRI shows that some problems in the satellite
attitude reconstruction affected this observation. The source appears
elongated in a way inconsistent with the expected Point Spread Function
(PSF). We note that this elongation ($\sim 20"$, approximately in
the north-south direction) is much greater than that occasionally
reported for other HRI sources (David et al. 1992). Clear displacements
in the source position can be seen by inspecting the images corresponding
to the individual time intervals which compose the observation. We
corrected the HRI image by dividing it in very short time intervals (20
s) and applying appropriate shifts to a reference position, in a way
similar to that described by Mereghetti et al. (1995). In the resulting
image the source profile is consistent with that of the instrument PSF
and the statistical significance of the weaker sources in the field has
increased. However, the derived absolute coordinates are clearly affected
by a systematic uncertainty of at least $20"$, since we had to
arbitrarily choose the reference position.
A boresight correction for the HRI data similar to that done for the PSPC
data is presently unfeasible, due to the ambiguos identification of the
HRI sources. We note that the GSC star used by
Walter et al. (1996) is at an off-axis angle of $21'$, where the
uncertainties in the HRI PSF largely affect the source location accuracy.

\bigskip
{\bf Table 1.} Photometry of the stars in the field of Fig. 1.
\medskip
\begintable
Star    \| $g$    \| $g - r$ \elt
 1      \| 21.77  \| 0.27 \el
 2      \| 18.30  \| 0.63 \el
 3      \| 21.50  \| 0.58 \el
 4      \| 19.40  \| 0.45 \el
 5      \| 21.80  \| 0.39 \el
 6      \| $>$22.7\| $>$0.6 \el
 7      \| 21.80  \| 0.81 \el
 8      \| 21.13  \| 0.54 \el
 9      \| 21.92  \| 0.96 \el
10      \| $>$22.7\| $>$1.15 \el
11      \| $>$22.7\| $>$0.6 \el
12      \| 21.81  \| 0.77 \el
13      \| 22.00  \| 0.04 \el
14 (C)  \| 18.27  \| --0.03 \el
15 (D)  \| 18.83  \| 0.33 \el
16      \| 20.15  \| 0.24 \el
17      \| $>$22.7\| $>$0.02 \el
18      \| 19.26  \| 0.99 \el
19      \| 22.50  \| 1.08 \el
20      \| $>$22.7\| $>$0.83 \el
21      \| 21.50  \| 1.32 \el
22 (F)  \| 17.54  \| --0.06 \el
23      \| $>$22.7\| $>$--0.12 \el
24      \| 20.89  \| 1.33 \el
25      \| 22.12  \| 0.17 \el
26      \| 21.64  \| 0.64 \el
27 (J)  \| 21.39  \| 0.85 \el
28      \| 21.85  \| 0.63 \el
29      \| 20.20  \| 0.40 \el
30 (I)  \| 19.09  \| 0.82 \el
31 (L)  \| 17.82  \| --0.18 \el
32      \| 18.80  \| 0.90 \el
33 (M)  \| 15.72  \| 0.44 \endtable
\medskip         

\section {Optical observations}

Optical images of the RXJ1856.5--3754 field were taken on 18--19 October
1995 with the Danish 1.54 m telescope of the European Southern Obervatory
(ESO) on La Silla. The telescope was equipped with the LORAL
$2052\times2052$ pixels CCD, with a pixel size of $0.39"$. Three exposures
of 5 minutes each were obtained in the $g$, $r$ (18 Oct.) and $i$ (19
Oct.) Gunn filters. 

Using the MIDAS/ROMAFOT photometric package, we have derived the $r$ and
$g$ magnitudes of several objects in the $\sim1'\times1'$ field around the
source position (see Table 1). No possible counterparts with particularly
unusual colours were found.

The deepest view through interstellar absorption was obtained in the
$i$ filter image (Figure 1), which however suffered from some
uncertainty in the absolute flux calibration. Several faint objects not
visible in the V band image of Walter et al. (1996) are clearly detected
inside the error circle. These sources are more clearly visible in the
contour plot shown in Figure 2. For instance, the star number 23 is
detected at a statistical significance of $\sim 4\,\sigma $ in both the
$i$ and $r$ band. We are therefore confident that they are real objects
and not background fluctuations.

\section{Spectral Analysis}

Source photons in the PSPC instrument were extracted from a circle of
radius $3'$ and the background from an annulus
centered on the source position of inner and outer radii of $3'$ and $6'$, 
respectively. Due to the very high count rate of
3.55$\pm$0.03 c s$^{-1}$, the fit results do not depend
significantly on the particular region used for the background
determination. The source is extremely soft, with very few counts
above $\sim1$ keV. We therefore rebinned the counts in the hard channels
in order to achieve a signal to noise ratio larger than 7 in each energy bin. 

Using XSPEC 9.0 software, we explored different
single component models. The best fit is obtained by a blackbody with a
temperature $T_{\rm bb}= 57.5^{+1.1}_{-2.7}$ eV and an absorbing column
density $N_H=(1.4^{+0.4}_{-0.2}) \times10^{20}$\cmdue. 
The unabsorbed 0.1--2.4 flux is to $3.8 \times10^{-11}$\ergs\cmdue.
The 0.1--2.4 keV luminosity derived from the black body model is
$1.3\times10^{32}\,R_6^2$\ergs\ (where $R_6$ is the neutron star radius
in units of 10 km), implying a distance of about $170\,R_6$ pc.
The relatively poor value of the reduced $\chi^2=1.8$ (42 d.o.f.,
probability $\sim 10^{-3}$) likely derives from calibration uncertainties in
the response matrix. These small uncertainties, which are particulary
evident in the region between 0.4 and 0.5 keV, are usually irrelevant in
the case of weaker sources, but dominate the fit residuals for such a
strong and soft X--ray source. By adding to the data a 5\% relative error
to account for this effect, the best fit $\chi^2$ drops to an acceptable
value of 1.1 (this additional systematic error is included in all the
following fit results). The best fit values of other simple models are
summarized in Table 2. 
         
\setcounter{table}{1}
\begin{table*}
\label{spe}
\stablesthinline=0pt
\stablesborderthintrue
\stablestyle=0
\caption{Summary of spectral fits for RXJ1856.5--3754 (errors are 90\%).}
\begintable
Model          | Column density      | Parameter                           |Red. $\chi^2$\el
Black body     |$1.4^{+0.4}_{-0.2}\times10^{20}$ cm$^{-2}$ |$T_{\rm bb}=57.7^{+2.5}_{-6.5}$ eV   | 1.1 \el
Bremsstrahlung |$2.3^{+0.4}_{-0.3}\times10^{20}$ cm$^{-2}$ |$T_{\rm br}=94.8^{+8.2}_{-6.2}$ eV   | 1.2 \el
Power law      |$4.1^{+0.5}_{-0.5}\times10^{20}$ cm$^{-2}$ |$\alpha=6.6^{+0.3}_{-0.3}$           | 1.9 \el
Raymond-Smith  |$5.5^{+0.5}_{-1.0}\times10^{20}$ cm$^{-2}$ |$T_{\rm RS}=53.2^{+3.9}_{-6.3}$ eV   | 1.7 \el
Zampieri       |$2.4^{+0.3}_{-0.3}\times10^{20}$ cm$^{-2}$ |$\log{L/L_{Edd}}=-8.47^{+0.29}_{-0.08}$| 1.2\endtable
\end{table*}
 
A detailed numerical analysis of the spectral properties of unmagnetized
neutron stars accreting well below the Eddington limit has been presented
by Zampieri et al. (1995). The emergent spectrum turns out to be
significantly harder than a blackbody at the star effective temperature.
Using this model, we obtained a reduced $\chi^2$=1.2 (42 d.o.f.; see
Table 2), corresponding to a total luminosity released by the neutron star
of $(4.3^{+4.0}_{-0.8})\times10^{29}$\ergs\, and a column density of
$N_H=(2.4^{+0.3}_{-0.3})\times10^{20}$\cmdue. The low luminosity implies
a very small distance of $\sim 10$ pc, inconsistent with the high column
density. 
These problems might also indicate that, contrary to the Zampieri et 
al. assumptions, a non-negligible magnetic field be present.
Nelson et al. (1995) have suggested that if ONSs have remained strongly
magnetized ($\sim 10^{12}$ G) there should be a cyclotron emission
feature in the hard X--ray band at $\sim 10$ keV on top of the soft
X--ray continuum, releasing up to 10\% of the total X--ray luminosity.
Observations at energies higher than the PSPC range can help
investigating this topic.
               
\section{Timing analysis}

The PSPC observation is sufficiently long to allow a periodicity search.
We take the same PSPC counts used for the spectral analysis, after
correcting their arrival times to the solar system barycenter. 
No periodicities were found. Assuming a sinusoidal light curve, we can put a 
$3\,\sigma$ upper limit of $\sim 15\%$ on the amplitude of the X--ray 
modulation in the range 0.1--10 s.

We extracted about 9700 counts from a circle of $80''$ radius around the source 
position in the HRI observation. No periodicities were found;
we can put a $3\,\sigma$ upper limit of $\sim 25\%$ in the range 0.04--100 s.

\section{Discussion}

We have shown that several optical objects are positionally compatible with
the revised error region of RXJ1856.5--3754. These possible counterparts
should be investigated in more details before looking for fainter objects.
In any case, also assuming a conservative error budget on the source
position, counterparts are limited to objects fainter than
$g\sim 18$ (F star in Figure 1). The X--ray to optical flux ratio of
RXJ1856.5--3754 is therefore $\gsim 20$, ruling out most classes of 
known X--ray emitters.

RXJ1856.5--3754 lies in the direction of the molecular cloud RCrA, whose
distance was estimated as $\sim 130$ pc (Dame et al. 1987). On the
basis of the low $N_H$ value derived from the PSPC fit, Walter et al.
(1996) concluded that RXJ1856.5--3754 is probably closer than the cloud.
However, RXJ1856.5--3754 does not coincide with the dense core of RCrA,
but it lies on the boundary of the cloud, in a region of much lower
extinction. Rossano (1978) derived an optical absorption $A_V\sim 0.7$ mag 
in the direction of RXJ1856.5--3754, compared with
$A_V \gsim 3$ mag at the cloud core. Also OH and HI maps
lead to a similar conclusion (Cappa de Nicolau \&
Poppel 1991). So it is not obvious that the source lies in front on the
cloud, based only on absorption considerations. 

This is in agreement with the distance estimated from the black body
spectrum ($170\,R_6$ pc) which places the neutron star beyond the
molecular cloud. At this distance the interstellar medium density is
about 1\cmtre, so that the neutron star velocity should be of $\sim 5$
km$\,$s$^{-1}$ in order to produce the observed luminosity. 
Taking the ONS velocity distribution of Blaes \& Madau (1993), we expect
0.15 \% neutron stars at such low velocities (1\% are slower than 10
km$\,$s$^{-1}$). Assuming a spatial density of $10^{-3}$ ONS 
pc$^{-3}$ we expect about 14 such slow objects within 170 pc (90
with $v < 10$ km$\,$s$^{-1}$). Therefore, even if different mechanisms
conspire against low velocity ONSs (Blaes, Warren \& Madau
1995), the derived numbers are large enough to retain the case of an ONS
 accreting directly from the interstellar medium. 

A different possibility is that RXJ1856.5--3754 is a young object still
dissipating its internal heat or emitting by some non-thermal processes. In
this respect, we note that different young ``radio quiet" neutron stars
are characterized by very soft spectra such as Geminga, and a few other
X--ray emitting neutron stars have recently been found associated to
supernova remnants.

\begin{acknowledgements}
We thank E. Molinari for taking the optical images and L. Zampieri 
for providing the spectral model.
\end{acknowledgements}

\end{document}